\documentclass[%
reprint,
showpacs,preprintnumbers,
 amsmath,amssymb,
aip,
 graphicx,
apl,
]{revtex4-1}

\usepackage{graphicx}
\usepackage{dcolumn}
\usepackage{bm}
\usepackage[mathlines]{lineno}

\begin{document}
\preprint{}

\title{Sub-nanometre resolution of atomic motion during electronic excitation in phase-change materials}

\author{Kirill V. Mitrofanov}
\author{Paul Fons}\email{paul-fons@aist.go.jp}
\author{Kotaro Makino}
\affiliation{Nanoelectronics Research Institute, AIST, Tsukuba Central 4, Higashi 1-1-1,Tsukuba 305-8562, Japan}
\author{Ryo Terashima}
\affiliation{Division of Applied Physics, Faculty of Pure and Applied Sciences, University of Tsukuba, 1-1-1 Tennodai, Tsukuba 305-8573, Japan }
\author{Toru Shimada}
\affiliation{Department of Science, Faculty of Education, Hirosaki University, 1 Bunkyo-cho, Hirosaki, Aomori 036-8560, Japan}
\author{Alexander V. Kolobov}
\author{Junji Tominaga}
\affiliation{Nanoelectronics Research Institute, AIST, Tsukuba Central 4, Higashi 1-1-1,Tsukuba 305-8562, Japan}
\author{Valeria Bragaglia}
\author{Alessandro Giussani}
\author{Raffaella Calarco}
\affiliation{Paul-Drude-Institut fur Festk\"orperelektronik, Hausvogteiplatz 5-7, 10117 Berlin, Germany}
\author{Henning Riechert}
\author{Takahiro Sato}
\affiliation{RIKEN SPring-8 Center, 1-1-1 Kouto, Sayo-cho, Hyogo 679-5148, Japan}
\author{Tetsuo Katayama}
\author{Kanade Ogawa}
\author{Tadashi Togashi}
\affiliation{Japan Synchrotron Radiation Research Institute, 1-1-1 Kouto, Sayo-cho, Hyogo 679-5198, Japan}
\author{Makina Yabashi}
\affiliation{RIKEN SPring-8 Center, 1-1-1 Kouto, Sayo-cho, Hyogo 679-5148, Japan}
\author{Simon Wall}
\affiliation{ICFO - Institut de Ci\`encies Fot\`oniques, The Barcelona Institute of Science and Technology, 08860, Castelldefels, Barcelona, Spain}
\author{Dale Brewe}
\affiliation{X-ray Science Division, Argonne National Laboratory, 9700 S. Cass Ave, Lemont, IL 60439, USA}
\author{Muneaki Hase}\email{mhase@bk.tsukuba.ac.jp}
\affiliation{Division of Applied Physics, Faculty of Pure and Applied Sciences, University of Tsukuba, 1-1-1 Tennodai, Tsukuba 305-8573, Japan }


\begin{abstract}
Phase-change materials based on Ge-Sb-Te alloys are widely used in industrial applications such as nonvolatile memories, but reaction pathways for crystalline-to-amorphous phase-change on picosecond timescales remain unknown. Femtosecond laser excitation and an ultrashort x-ray probe is used to show the temporal separation of electronic and thermal effects in a long-lived ($>$100 ps) transient metastable state of Ge$_{2}$Sb$_{2}$Te$_{5}$ with muted interatomic interaction induced by a weakening of resonant bonding. Due to a specific electronic state, the lattice undergoes a reversible nondestructive modification over a nanoscale region, remaining cold for 4 ps. An independent time-resolved x-ray absorption fine structure experiment confirms the existence of an intermediate state with disordered bonds. This newly unveiled effect allows the utilization of non-thermal ultra-fast pathways enabling artificial manipulation of the switching process, ultimately leading to a redefined speed limit, and improved energy efficiency and reliability of phase-change memory technologies.
\end{abstract}

\flushbottom
\maketitle

\thispagestyle{empty}

\section*{Introduction}

The class of chalcogenide phase-change materials, such as Ge-Sb-Te and Ag-In-Sb-Te, has been found to be the most appropriate candidate for optical data storage media in the forms of rewritable CDs and DVDs, as well as Blu-ray discs. Ge$_{2}$Sb$_{2}$Te$_{5}$ (GST) has also been demonstrated to be a compound well suited for non-volatile memory applications, owing to a fast phase-switching process between amorphous and crystalline phases (10 $\sim$ 100 ns), excellent thermal and chemical stability of the end phases and good reliability allowing more than 10$^{9}$ write-erase cycles\cite{Cheng2013}. GST superlattice films are now being applied in a new generation of non-volatile electrical memory, interfacial phase-change memory\cite{Simpson2011nn}, surpassing current FLASH technology both in cyclability and speed\cite{RaouxPCM2009}. The process of rapid phase change involved in the writing and erasing of data in conventional optical recording is presently induced by a purely thermal process using nanosecond laser pulses: heating of the material leads to the formation of a molten phase and subsequently the crystalline (SET) or amorphous (RESET) state, depending on the cooling speed. Thus in phase-change memory devices the speed of write cycles is limited to tens of nanoseconds\cite{Wuttig2007nmat}. Therefore in order to overcome this limitation, an alternative method of switching between SET and RESET states of GST is necessary. Avoidance of thermally based processes implies the necessity of using electronic effects, and thus the nature of the electronic structure of GST needs to be taken into account. While its amorphous phase is characterized by its covalent bonding nature, the distorted rock-salt crystalline phase is locally rhombohedral with three long and three short bonds and is usually described in terms of resonant bonds\cite{shportko2008nm,Kolobov2012pss,krbal2012prb}. Recently, the possibility of an ultrafast phase transition triggered by an electronic excitation due to the breaking of resonant bonds has been proposed\cite{kolobov2014jpc,Li2011prl}. Based upon the experimental results from a time-resolved x-ray absorption fine structure measurement, the possible presence of non-thermal
contributions to the amorphization of GST alloy on sub-nanosecond time scales was reported \cite{Fons2010prb}. The existence of a solid-solid amorphization process induced via electronic excitation and subsequent lattice relaxation was further argued for by density functional calculations \cite{Kolobov2011nc} and time-resolved
electron diffraction studies\cite{Hada2015,Hu2015}, in which, however, due to the polycrystalline samples used and the transmission mode applied, the direct demonstration of the existence of electronically driven effects without the significant influence of thermal processes is challenging. This reason, in addition to the fact that electronic excitation is in general immediately followed by electron-phonon coupling-induced heating of the lattice, makes the dynamics induced by an increase in lattice temperature difficult to separate from purely electronic effects.

Very recently, Waldecker {\it et al.} reported on the decoupling of the electronic and lattice degrees of freedom on a several picosecond time scale after optical excitation. They interpreted the temporal separation of optical properties from the structural transition in GST alloy films to be a result of the depopulation of resonant bonds before electron-phonon energy exchange occurred\cite{Waldecker2015}. These observations on a several picosecond time scale will open a new route to control phase change in resonantly bonded materials. However, the observation of the decoupling was limited to a narrow time window, preventing from evaluating the lifetime and exact structure, both of which are required for real device applications. Here we provide new insight into a possible solid-solid transformation process in an epitaxially-grown single-crystal film of GST based on reversible conditions through time-resolved x-ray techniques with sub-nanometre resolution allowing the temporal ranges of the lattice evolution to be discerned. The use of a much wider time window up to $>$ 1 ns reveals the long lifetime of the transient state and associated bond angles disordering. The x-ray diffraction intensity decreases as a function of time delay without a thermal shift in peak position for several picoseconds after photo-excitation, followed by the sudden onset of a thermal-expansion-induced shift. These features coincide well with previous results\cite{Waldecker2015}, but, being obtained from the direct observation of atomic motion as opposed to indirect information derived from the evolution of the optically excited dielectric function, strongly and unambiguously support the existence of the separation of non-thermal (electronic) effects from thermal (lattice) effects. These observations were also made possible by the use of an epitaxial sample in the rocking curve mode measurements, providing clear evidence of the movement of atoms independent from the optically-induced heating of the sample. Our x-ray absorption fine structure (XAFS) data further prove that the transient state is an intermediate state between the original crystalline and amorphous states, but different from the liquid phase that would be expected to exist for the case of switching based on a melting process.

\begin{figure}[ht!]
\centering
\includegraphics[width=88mm]{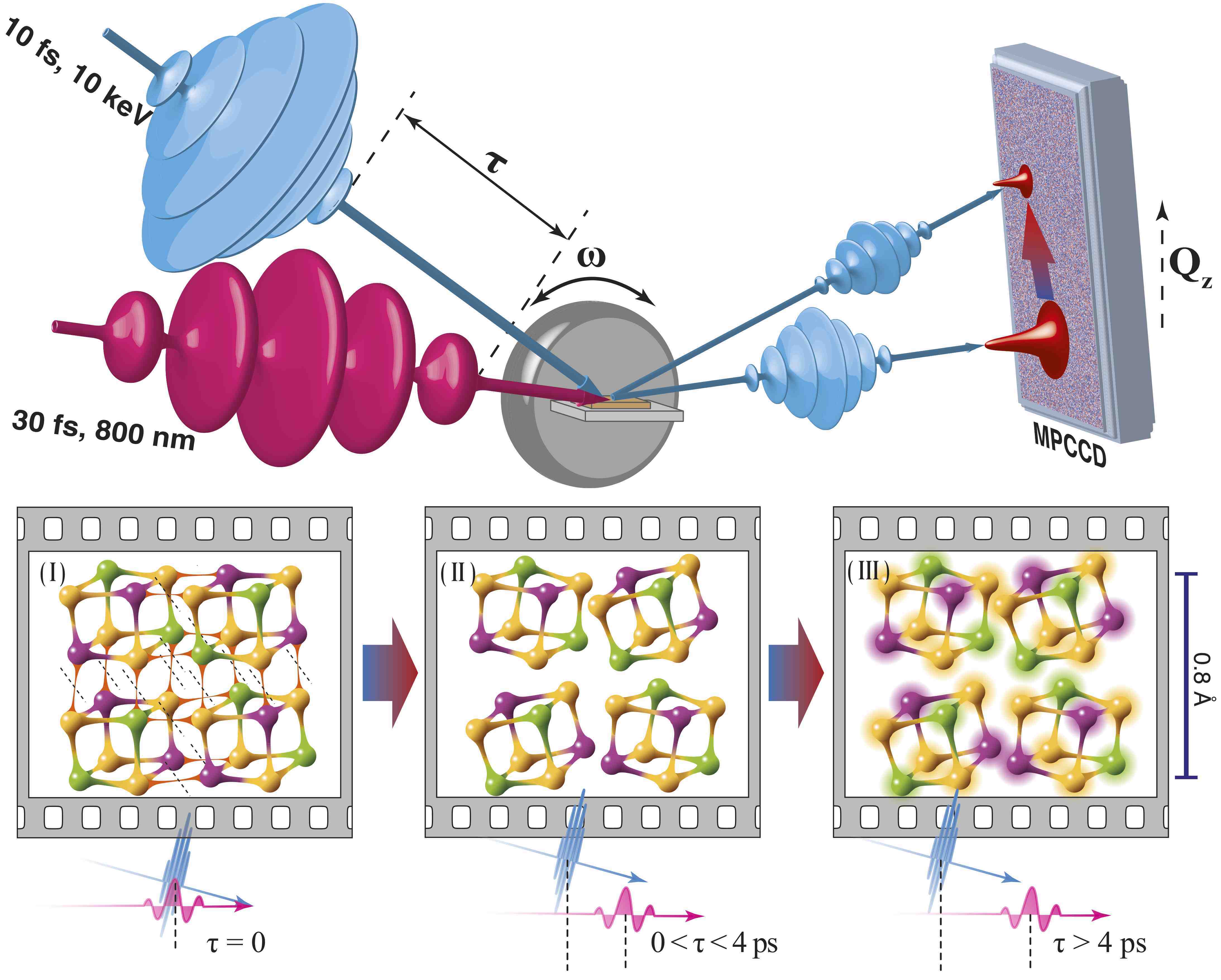}
\caption{\textbf{Layout of the experiment and schematic representation of the GST structural change.} The sub-nanometre dynamical atomic configuration processes in GST are schematically shown at the bottom with three different time frames. (I) Before the excitation ($\tau = 0$) the atomic configuration is resonantly bonded (resonant bonds are shown in red color thin interconnections between atoms). (II) At $0 < \tau < 4$ ps the resonant bonds are disrupted and local lattice distortions appear, while no thermal energy from the electrons was transferred to the lattice, which corresponds to a new metastable state. (III) At $\tau > 4$ ps the optical phonons start to emit (heating of the lattice), giving rise to bigger oscillations amplitudes of the atoms, which is represented by thermal ellipsoids around them.}
\end{figure}

\section*{Results}

We used a time-resolved x-ray diffraction (XRD) technique\cite{Siders1999,Sokolowski-Tinten2003nat,Ichikawa2011} (Figure 1) to directly probe the ultrafast structural dynamics in a 35-nm-thick single crystalline GST film grown on a Si (111) substrate photo-excited by 1.55 eV (800 nm) photons. The laser-induced dynamics of the GST lattice were studied by taking time-resolved XRD rocking curves of the GST (222) Bragg diffraction peak (Figure 2a,b). This peak served as an optimal peak for observations of the lattice dynamics due to the boundary conditions imposed by the pump laser; the corresponding integrated intensities and peak shift were measured as a function of the time delay (Figure 2c,d). The diffraction \textit{intensity} of the (222) peak decreases immediately after excitation, and at $\sim$4 ps the intensity level reaches $\approx$ 80\% of its initial value. In contrast, the (222) diffraction peak \textit{position} starts to shift towards lower $Q_{z}$ values only after a $\sim$4 ps time delay with a concomitant further decrease in intensity (Figure 2d).  The integrated intensity taken at a fixed incident angle at the (222) peak position with high time resolution ($Q_{z}$=const) shows that the diffracted signal intensity has an inflection point at $\approx$ 4 ps, which, together with a combination of intensity and position of the diffraction peak, demonstrates the existence of two-step dynamics. The (222) peak reaches a maximum deflection $\sim$ 20 ps after the excitation. The shift of the position of the Bragg peak towards smaller $Q_{z}$ values after excitation implies that the maximum thermal expansion of the GST film is \textit{${\Delta}$}\textit{d}/\textit{d} $\approx$ 1.4\%\cite{Fons2014}, where $\Delta d$ is the fractional change in $d$ spacing. Subsequently, the peak position reverts to higher $Q_{z}$ values, corresponding to contraction of the lattice.  

The unshifted position of the (222) diffraction peak, observed with better time resolution than in previous work\cite{Fons2014}, implies that thermal effects leading to expansion of the lattice do not occur during the first $\sim$4 ps, since a diffraction peak shift toward lower $Q_{z}$ values would be expected if the lattice temperature had risen due to electron-phonon interactions\cite{Lahme2014}.  Such a delay in the lattice heating onset may be due to the presence of charge screening leading to a decrease of electron-phonon coupling and/or Auger recombination in the early stages of the carrier relaxation process\cite{Cavalleri2000,Downer1986,Chin1999}. These processes make possible the conservation of electron energy, leaving the electrons excited for a time much longer than the characteristic time of optical phonon emission. Such a long-lived excited state in GST results in a specific non-thermal lattice response in which nanoscale local order changes due to the high concentration of electrons remaining in an excited state. This premise is supported by a combination of the diffraction main peak intensity and position dynamics: while the intensity of the diffraction signal has a contribution from thermal effects, including the Debye-Waller factor (DWF), the diffraction peak position reflects solely the lattice strain (expansion or contraction) along the direction of the scattering vector. The correlation between the increase in the lattice temperature and the shift of the diffraction intensity peak is based on the fact that the excess energy transferred to the lattice from the optically excited electrons leads to an increase in the stress, which has gradients at the film interfaces. This produces a strain wave propagating into the sample at the speed of sound and an increase in the portion of the film with modified lattice spacing as the wave propagates. This results in a new Bragg condition and changes in the corresponding dynamics of the diffraction rocking curve profile, in particular, a shift of the main diffraction peak position.

\begin{figure}[ht!]
\centering
\includegraphics[width=88mm]{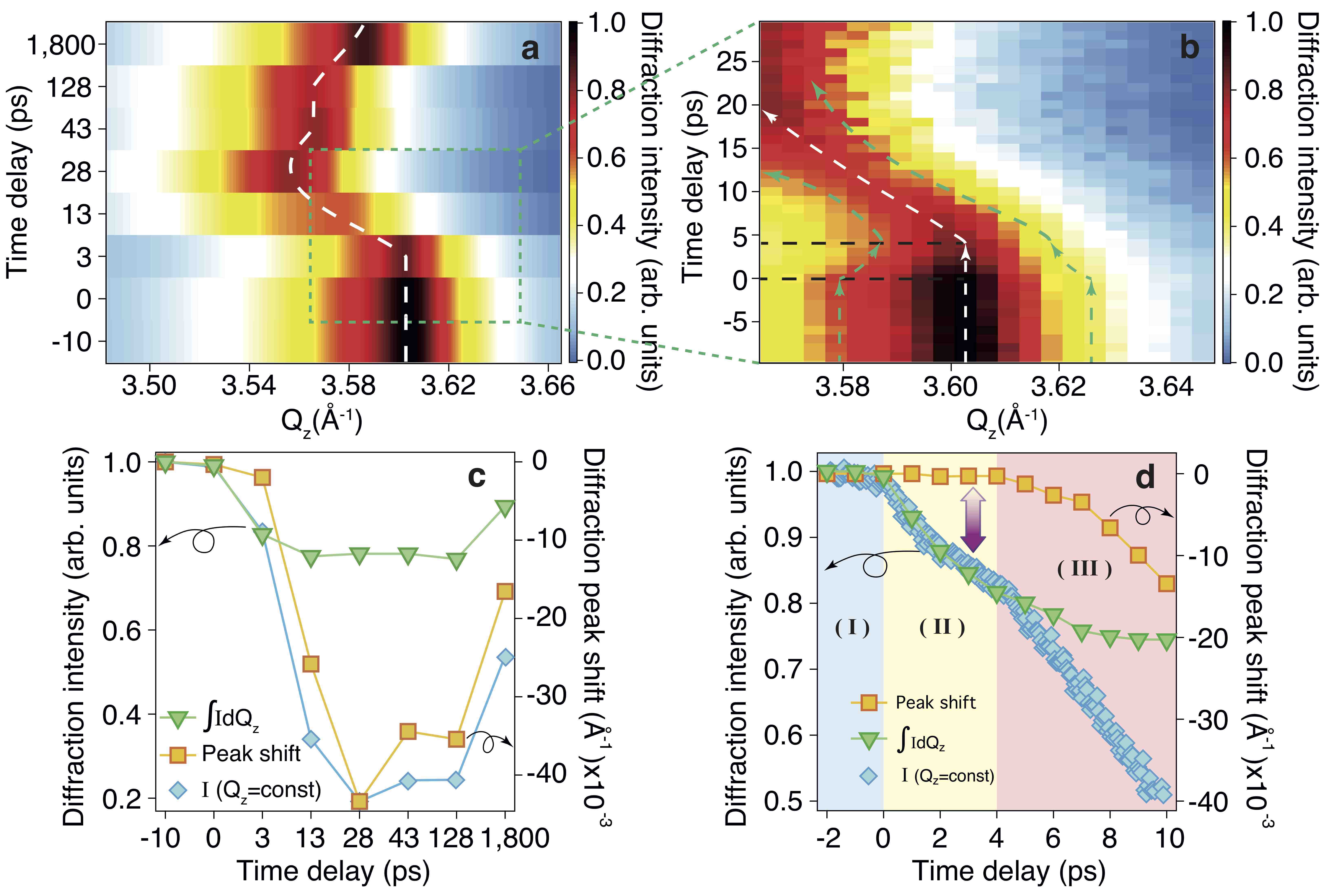}
\caption{\textbf{Time-resolved x-ray diffraction results. (a)} X-ray diffraction rocking curves as a function of time delay for the GST (222) reflection centered at 3.603 \AA$^{-1}$, observed for a 13.9 mJ cm$^{-2}$ pump fluence and a $\geq$10 ps time delay step, over a wide $Q_{z}$ range, where $Q$ represents the momentum transfer of the elastic scattering process and $z$ is the direction of the scattering vector. The white dashed line traces the center of the diffraction peak. \textbf{(b)} The same as \textbf{(a)} but with a 1 ps time delay step, over a narrow $Q_{z}$ range. The green dashed lines trace the diffraction intensity level. The black dashed lines indicate  time zero and the time delay corresponding to the beginning of the diffraction peak shift. \textbf{(c)} Normalized integrated diffraction intensity (inverted triangles) and corresponding peak position shift along $Q_{z}$ (squares), together with normalized integrated diffraction intensity (diamonds) for a fixed x-ray beam angle of incidence as a function of time up to 1.8 ns. \textbf{(d)} The same as \textbf{(c)} but for a finer time delay step up to 10 ps. The arrow indicates the different dynamics between the diffraction integrated intensity and the peak position. The data evolution is divided into three stages using background color of the plot, corresponding to the atomic configuration process in Figure 1: ground state - (I), disrupted crystalline state with a cold lattice - (II) and a state with thermal effects started to take place - (III).}
\end{figure}

\section*{Discussion}

Moderate decreases in diffraction intensity can be described by the Debye-Waller factor, $e^{-\frac{1}{2}Q^{2}\langle u^{2}\rangle}$, where $\langle u^{2}\rangle$ denotes the mean-square displacement of collective atoms along the scattering vector $Q$\cite{AlsNielsenXrayPhys2011}, but the absence of a thermally-induced diffraction peak position shift in the experimental data is an evidence of a different scenario. While one might assume that the observed changes are result of a response from a complete disordering of the part of the crystal lattice, this explanation is not valid: dramatic decreases in the diffraction intensity from crystalline samples are usually associated with structural disorder, which in turn is connected with randomization or fluctuations in atomic positions, but in tetrahedral covalently bonded semiconductors non-thermal melting has been reported to occur within several hundred femtoseconds\cite{lindenberg2005sci}, a time scale much faster than seen in the current GST film. In fact, the observed recovery of the Bragg peak intensity on the nanosecond time scale demonstrates that the initial crystalline state remains. Re-crystallization from the molten state also cannot explain the observed dynamics as optically induced crystallization processes in GST require much longer times - typically more than 10 ns\cite{Siegel2004}. Therefore, the observed decrease in diffraction intensity for times $<$ 4 ps cannot be interpreted to be the result of non-thermal melting in the general sense, i.e. a complete loss of long-range order. The absence of a thermal-shift during the first 4 ps, however, indicates the presence of non-thermal effects in photo-excited GST, arising due to the effects of intense optical excitation on resonant bonding. It should be noted that the term "resonant bonding" with respect to phase-change materials is used differently by different authors: while some authors consider all bonds to be resonant, others refer to only longer bonds as being so. In this manuscript, we have adopted the latter approach.
The experimentally observed dynamics indicate that the diffraction peak intensity does not fully recover even 1.8 ns after excitation (Figure 2c) due to residual thermal effects, which completely disappear after 3 $\sim$ 5 ns depending on the excitation fluence. The corresponding return of the diffraction peak to its initial intensity demonstrates that the GST film was not amorphized, but transiently transformed to an intermediate state. The drop in the diffraction signal intensity cannot be explained soley by the DWF for the following two major reasons. First, the absence of the diffraction peak position shift until 4 ps after the excitation, which is more clearly represented by an inflection point on the time-dependent diffraction intensity curve, indicates the lack of thermal effects immediately after laser exposure. The inflection point marks the start of the shift of the diffraction signal peak position due to the expansion of the GST film, i.e. the onset of thermal effects. Second, if one assumes that the diffraction peak shift is the result of the strain in the sample due to thermal effects, then the acoustic phonon-induced contribution of the strain, corresponding to the shift at 128 ps after excitation, should be highly damped. This assumption leads to a calculated lattice temperature using the thermal expansion coefficient of 1.74x10$^{-5}$ K$^{-1}$ (ref. \cite{Park2008tsf}) of $\approx$ 930 K, a temperature which is higher than the melting point, which is inconsistent with observation and thus indicates the presence of long-lived non-thermal effects\cite{shu2013apl}.

\begin{figure}[ht!]
\centering
\includegraphics[width=88mm]{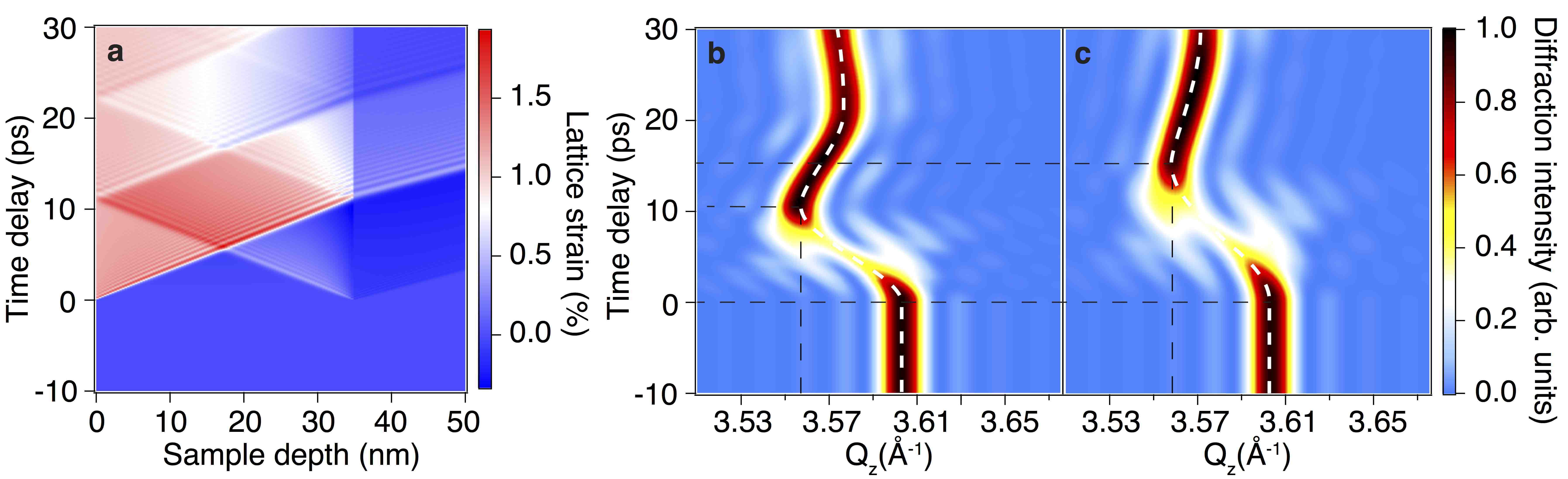}
\caption{\textbf{Simulation results for the lattice dynamics of an optically excited GST sample. (a)} Time dependence of the crystal lattice strain for the case of the current experiment conditions except the literature value of the sound velocity (3.19 nm ps$^{-1}$)(see ref. \cite{Fons2014}). \textbf{(b)} Corresponding diffraction data evolution with time for the (222) reflection, the white dashed line traces the diffraction peak position, the black dashed lines indicate the value of the diffraction maximum peak shift along $Q_{z}$ (vertical line) and time delays corresponding to the beginning of the peak shift and the maximum shift (horizontal lines). \textbf{(c)} The same as \textbf{(b)}, but for the case of the sound velocity value obtained from the current experiment (2.19 nm ps$^{-1}$).}
\end{figure}
In addition to thermal expansion, the laser-induced thermal stress in the film gives rise to the generation of acoustic phonons originating from the interfaces of the film with the capping layer and the substrate (Figure 3)\cite{Thomsen1986}. The half-period of the acoustic phonon \textit{(T}/2 is equal to the ratio of the sample thickness \textit{(l)} and the sound velocity \textit{(v)}: \textit{T}/2=\textit{l}/\textit{v}. The half-period of the experimentally observed strain waves originating from the interfaces was estimated to be $\approx$ 16 ps (Figure 3b). This allows the calculation of the sound velocity in the metastable state to be $\approx$ 2.19 nm ps$^{-1}$, a value which is $\approx$ 2/3 of the literature value for crystalline GST\cite{Fons2014,lyeo2006apl} and close to the value for the amorphous case\cite{lyeo2006apl}.
To account for the observed strain dynamics, the propagation of strain waves in laser-excited GST as well as the resulting transient diffraction intensity change was simulated using a one-dimensional linear chain model as implemented in the package UDKM1DSIM\cite{Fons2014,Schick2013cpc} for two cases: using the literature value of the sound velocity for crystalline GST and the experimentally estimated sound velocity as an input parameter. In the first case the obtained time delay for the maximum deflection of the diffraction peak was about 2/3 of the half-period of the experimentally observed strain wave (Figure 3b). On the other hand, for the case of the experimentally estimated speed of sound value, the simulation results are in good agreement with the experimental data (Figure 3c), which indicates a significant change in the interatomic potentials from the ground state, a result which can be related to the local reorientation of atoms ensembles, leading to increased bonds lengths and a wider angular distribution of bonds.
The results of the simulated transient diffraction pattern obtained using the UDKM1DSIM package are in a good agreement with the experimental data for the case of the experimentally estimated speed of sound in the GST film, but it should be noted that an earlier work, in which much slower optical excitation was used, the simulation data matched the experimental results for the literature value of the sound velocity\cite{Siegel2004}. This disparity is a consequence of the differences in excitation conditions (in the current work the pump pulse is much faster: 30 fs instead of the 700 fs (presented in ref. \cite{Fons2014}), giving rise to the markedly different dynamics observed in the current work. In addition it should be pointed out that the speed of the diffraction response in the earlier work was also masked by a $\approx$100 ps convolution in time, due to the probe pulse width, making small changes in sound velocity difficult to distinguish.

We argue that the ultrafast excitation of electrons from bonding to anti-bonding states leads to the breaking of longer (weaker) Ge/Sb-Te bonds\cite{kolobov2014jpc,Waldecker2015}, i.e., the non-thermal melting of resonant bonds, leaving stronger covalent bonds intact. Based on the results presented in refs. \cite{Lee2011prl} and \cite{Akola2007spt}, we schematically represent the excited structure as non-connected cubes (Figure 1). The process of the formation of such building blocks is expected to occur over very short times, less than the characteristic recombination time of non-equilibrium charge carriers. The breaking of the longer Ge/Sb-Te bonds results in local structural relaxation, while the average long range order in GST crystal persists. The corresponding state can be characterized as a "disrupted crystalline" state. This transient disordering in GST is schematically visualized in Figure 1. In the ground state, building blocks are connected to each other by resonant bonds; in contrast, upon electronic excitation the building blocks locally relax due to the loss of resonant bonding, leading to the tilting and shifting of the blocks from their initial positions\cite{Kolobov2011nc,kolobov2014jpc}. 

The decrease in the (222) Bragg diffraction peak intensity continues until the excited system gradually reverts to its initial state with resonant bond interaction over the course of at least 100 ps (Figure 2c). This implies that despite in the initial stage (first 4 ps) the excited state is purely electronic, it persists for much longer times, accompanied and thus hidden by thermal effects, which does not prevent its usage for ultrafast structural modification. Upon photo-excitation, local distortions can lead to a change in some Ge atomic configurations from an "octahedral" to a tetrahedral geometry with \textit{sp$^{3}$}-hybridization\cite{Akola2007spt,caravati2007cta,Micoulaut2014}. Local structural relaxation back to the ground state is suppressed by a decrease in the probability of the restoration of resonant bonding due to a transient increase in mean-square atomic displacements and fluctuations in bond angles. These effects extend the long relaxation period of the excited electrons and serve as a kinetic barrier leading to the unusually long duration of structural recovery from the excited state in GST.  

The existence of a metastable (intermediate) state is also supported by an independent time-resolved XAFS experiment. A unique XAFS  $k^2 \chi(k)$ signal was collected in which an excited state structure of GST (Figure 4a), different from both the liquid and amorphous states (Figure 4b), was transiently observed upon ultrafast optical excitation with the sample quickly reverting to its crystalline state (within $\approx$1 ns). 
\begin{figure}[ht!]
\centering
\includegraphics[width=60mm]{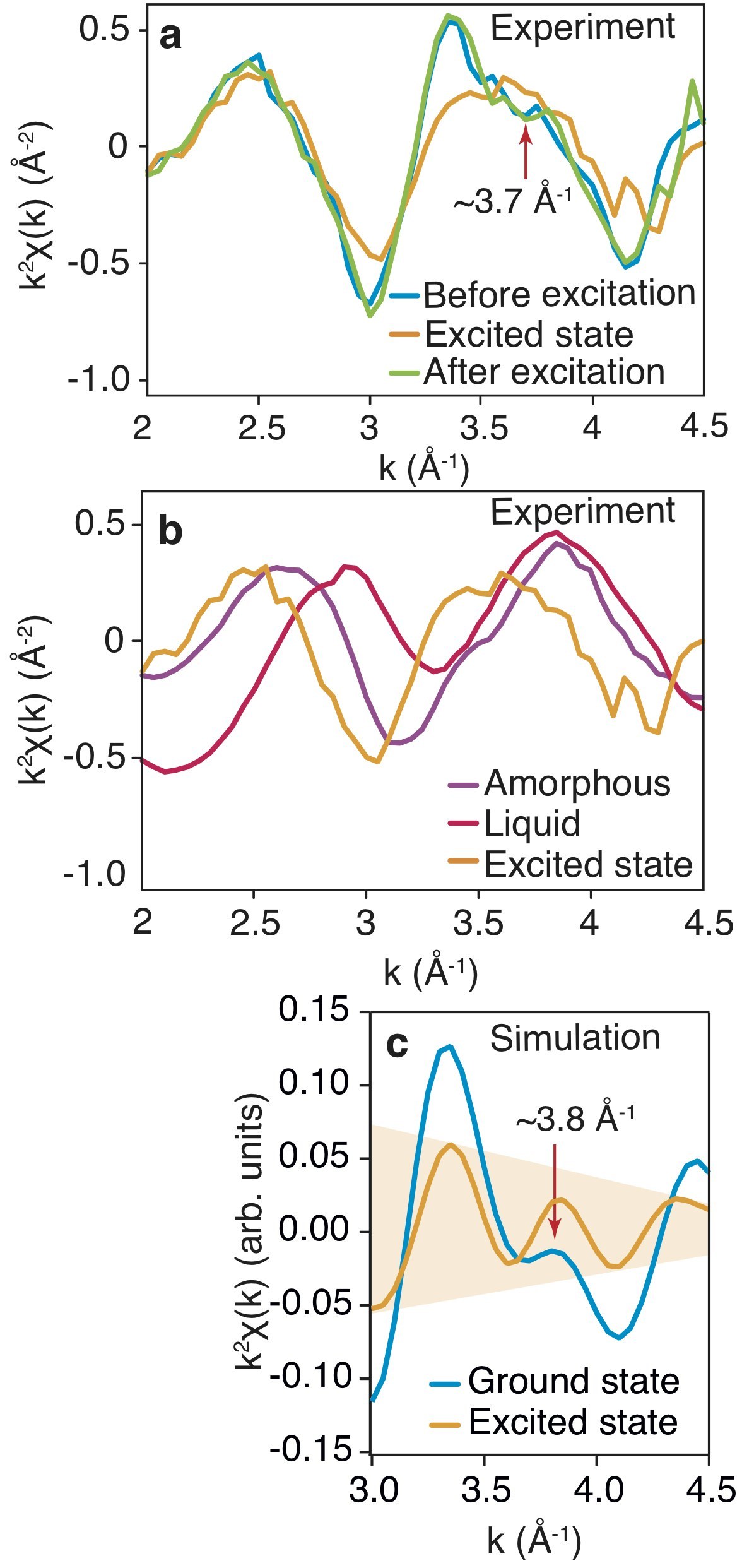}
\caption{\textbf{XAFS experimental and simulations results. (a)} Ge K-edge XAFS oscillations of GST for the crystalline cubic state before, during and after the excitation, the red arrow points to the "beat" in the ground state and the state after the excitation at $\sim$3.7 \AA$^{-1}$. \textbf{(b)} Ge K-edge XAFS oscillations of GST for amorphous, liquid and excited state. \textbf{(c)} MD simulation results of Ge K-edge XAFS signal for 3 $<k<$ 4.5 \AA$^{-1}$ (see explanations in Methods section) for GeTe in the ground and excited states, the arrow points out the "beat" in the ground state at $\sim$3.8 \AA$^{-1}$. The light-orange area in the plot traces the damping of the function, corresponding to the excited state.}
\end{figure}

A similar change could be reproduced based upon a molecular dynamics (MD) simulation using the plane wave density functional program VASP (see Experimental Section for details), in which 4\% of the valence electrons were placed into the conduction band (a value equal to the theoretically estimated concentration of the excited electrons for the experimental conditions used); the MD simulations were carried out using a Nos\'e thermostat, which maintained the lattice temperature at approximately 300 K (the standard deviation of the temperature fluctuations was approximately 20 degrees implying that temperature fluctuations do not lead to significant changes in the XAFS spectra and can be neglected), i.e. the observed change is a consequence of a purely electronic effect. Figure 4c shows the results of the simulated time averaged XAFS signal for both the ground and excited states.  In the MD simulations, we have used GeTe as an approximant for GST to avoid the configurational complexity of the vacancy distribution associated with GST.  While for the ground state spectra, obtained from the experiment, there is a clear "beat" at approximately 3.7 \AA$^{-1}$ that arises from quantum mechanical interference of the outgoing photoelectron wave function with itself due to backscattering from a well-defined second coordination shell, for the excited state, the interference is strongly diminished, and the spectra itself closely approximates a damped sine function, a shape characteristic of backscattering from a single coordination shell (Figure 4a). The simulations of the 300 K systems reproduce the experimental trends well with the difference between the two simulated spectra being solely due to the change in the electronic conditions, strongly supporting the non-thermal nature of the lattice perturbations. This is interpreted to be a consequence of ultrafast strong optical excitation leading to a transient increase in lattice local disorder;  the same disorder is speculated to lead to the experimentally observed transient dramatic change in sound velocity and the extraordinary behavior of the diffraction signal peak position/intensity, measured in the XFEL pump-probe experiment.

The existence of a disrupted crystalline state in GST can be used in phase-change memory to transiently lower the potential barrier between the SET and RESET states allowing energetically preferable switching with anticipated improvements over the current speed limit to rates in the GHz region (which corresponds to a few picoseconds for the structural relaxation to occur). The long lived excited state can be utilized in a different types of applications, like optical and electrical switches and sensors, possibly working even in the THz region, since the observed excited state appears in the first tens of femtoseconds (just after the laser excitation).

The process of breaking the resonant bonds, leading to significant lattice local distortions, can serve as a precursor of non-thermal high-speed phase-switching. Moreover, it represents a new metastable state, characterized by both electronic and crystal lattice conditions, that results in optical and electric properties, different from the ground state. This difference can be utilized for the introduction of fundamentally new concepts of device operation for the recording and processing of data, including the possibility for applications in multi-level memory devices. The understanding of the transformation dynamics on ultrafast time scales over a few picoseconds provides new insight into the ultimate speed limit of phase change materials\cite{Loke2012} and suggests that electronic effects may play a heretofore unrecognized important role in device switching, possibly leading to significant reductions in the energy requirements. Ultimately, resonant bonding state manipulation may allow the alteration of the crystal structure solely by the selective modification of the electronic state. In particular, the recent development of interfacial phase-change memory with spatially separated GeTe and Sb$_{2}$Te$_{3}$ layers, in which switching between the SET and RESET state is achieved without melting\cite{Simpson2011nn}, may be assisted by the presence of excited states induced by the large electric fields applied during switching. 

\section*{Methods}
\subsection*{Sample preparation:}
Crystalline Ge$_{2}$Sb$_{2}$Te$_{5}$ (GST) thin films were fabricated by molecular beam epitaxy (MBE) on the Si(111)-($\surd$3x$\surd$3)R30$^{\circ}$-Sb reconstructed surface as described in detail elsewhere\cite{Boschker2014nano}. According to x-ray reflectivity measurements (not shown) the film thickness of the sample was 35 nm. The GST films were investigated by x-ray diffraction (XRD) and showed a cubic phase characterized by a high degree of texture and structural order in both the out-of-plane and in-plane directions \cite{Boschker2014nano,Katmis2011,Rodenbach2012pss}. The lattice constant $a$ was 6.04 \AA{}, which is consistent with the literature value of 6.01 \AA{} $\sim$ 6.04 \AA{}\cite{Rodenbach2012pss}. Neglecting the slight rhombohedral distortion, GST is hence lattice-mismatched by $\approx$11\% to Si (\textit{a} = 5.431 \AA{}). The samples were capped with 30 nm of sputtered amorphous Si$_{3}$N$_{4}$ to prevent oxidation. 
\subsection*{Pump-probe x-ray diffraction:}
Ultrafast time-resolved XRD measurements were conducted at the Japanese XFEL facility SACLA\cite{Tono2013} using the output of the x-ray FEL in spontaneous emission mode. A monochromatic beam was generated from the XFEL beam using two channel-cut Si(111) crystals in a non-dispersive arrangement corresponding to a wavelength of 1.24 \AA{} with a FWHM pulse of $\approx$ 10 fs. This beam was directed onto the sample as a probe pulse and the diffraction reflections obtained in the Bragg condition were detected by a 512 x 1024 channel multi-port CCD camera (Figure 1). Each measurement was repeated up to 250 times to improve statistics. A high-speed chopper was used to isolate individual x-ray pulses at 30 Hz. The p-polarized optical laser pulses for the pump ($\approx$30 fs-duration at a center wavelength of 800 nm) were synchronized with the XFEL; the angle between the laser and x-ray beams was $\approx$5$^{\circ}$. The maximum pump power was 3 mW (or 55 $\mu$J pulse$^{-1}$) and the laser was focused into a 700 $\mu$m diameter spot, a size significantly larger than the diameter of x-ray beam spot (250 $\mu$m). The results presented here were obtained with pump fluences up to 13.9 mJ cm$^{-2}$ at the sample surface, which is below the threshold value for the irreversible changes in GST\cite{Waldecker2015} after taking into account the reflectivity and the ellipsoidal broadening of the laser beam spot for the used angle of incidence. The jitter in the time delay between x-ray and laser pulses was $\leq$ 0.2 ps, supporting an effective time resolution better than 200 fs. The time delay ($\tau$) between the pump laser and the x-ray pulse was adjusted by an optical delay line in time increments of 60 fs to 100 ps. An effective time resolution of 60 fs was obtained by time sorting x-ray pulses using a timing monitor for the diffraction intensity plot shown in Figure 2d.
\subsection*{Time-resolved XAFS measurement:}
XAFS measurements were taken at sector 20 of the advanced photon source (APS). In the experiment, a 40 nm thick polycrystalline layer of GST encapsulated by 20 nm thick transparent Si$_{3}$N$_{4}$ layers with a ZnS-SiO$_{2}$ protective cap was irradiated by a 190 fs, 800 nm pump laser pulse and the resulting lattice dynamics were probed via measurement of Ge K-edge XAFS as a function of laser delay. The fluence used was 9 mJ cm$^{-2}$. The spot size of the laser was approximately 55 $\mu$m and the $\sim$ 5 $\mu$m Kirkpatrick-Baez focused x-ray probe beam was well contained within the pump beam spot.  A quartz disk served as the substrate and was continuously rotated using an ultra-low wobble spindle motor to eliminate heat build up and reduce radiation damage. The laser system consisted of a Coherent Millenia Ti-Sapphire oscillator phase-locked to the RF signal of the storage ring and a Coherent RegA regenerative amplifier system which could provide power up to 4 $\mu$J pulse$^{-1}$. The disk was rotated assuring that several microseconds passed before a given area on the sample was subsequently irradiated.  The Ge fluorescence signal was detected via a 100,000 element gated Pilatus detector. By varying the laser trigger signal delay, the relative position of the laser pump to the x-ray probe pulse could be varied in $\sim$ 18 ps intervals. The x-ray probe pulse had a duration of approximately 100 ps resulting in a similar convolution in time of the experimental data. The fact that the unique signature of the excited state is visible despite the convolution resulting from the duration of the x-ray pulse, substantiates the long lifetime of the excited state.
\subsection*{VASP simulations:}
The simulated XAFS spectra were generated in two steps. In the first, ab-inito molecular dynamics (AIMD) were used to follow the trajectory of a 128 atom GeTe cluster for 30 picoseconds using density functional theory and the plane wave code VASP 5.4.1\cite{VASP}. The effects of optical excitation were simulated by constraining the band occupancies such that 4\% of the valence electrons were promoted to the conduction band. TThe generated trajectories were then used as input to the real-space multiple scattering code FEFF 9.64\cite{feff8} to generate XAFS spectra, which were then averaged to generate the final simulated XAFS spectra.  A cutoff energy of 175 eV was used for the plane waves in VASP and k-point sampling was carried out using the gamma point. A convergence study showed that use of the gamma point for integration of the Brillioun zone was sufficient to reflect the the general trends of the XAFS data. The energy difference was found to be less than 10 meV/atom. Projector augmented waves were used to include the effects of the core electrons. In Figure 4c the data is presented for k $>$ 3 \AA{}$^{-1}$, due to limitations of the extended x-ray absorption fine structure approximation used.

\section*{Acknowledgements}

This work was supported by X-ray Free Electron Laser Priority Strategy Program, entitled Lattice dynamics of phase change materials by time-resolved X-ray diffraction (NO. 12013011 and 12013023), from the Ministry of Education, Science, Sports, and Culture of Japan.

The XFEL experiments were performed at the BL3(EH2) of SACLA with the approval of the Japan Synchrotron Radiation Research Institute (JASRI) (Proposal No. 2012B8041, 2013A8051, 2013B8056, 2014A8039 and 2014B8061).

Sector 20 facilities at the Advanced Photon Source, and research at these facilities, are supported by the US Department of Energy - Basic Energy Sciences, the Canadian Light Source and its funding partners, the University of Washington, and the Advanced Photon Source. Use of the Advanced Photon Source, an Office of Science User Facility operated for the U.S. Department of Energy (DOE) Office of Science by Argonne National Laboratory, was supported by the U.S. DOE under Contract No. DE-AC02-06CH11357.

\section*{Author contributions statement}

K.V.M., P.F., K.M., R.T., T.Shimada, A.V.K., A.G., T.Sato, T.K., K.O., T.T., S.W., D.B. and M.H. carried out the experiments. K.V.M., P.F. and M.H. processed the data and performed the simulations. V.B., A.G. and R.C. prepared the samples. K.V.M., P.F., A.V.K. and M.H. wrote the manuscript. All of the co-authors contributed to the discussions of the results and manuscript.

\section*{Additional information}

\textbf{Competing financial interests:} 
The authors declare no competing financial interests. 


\begin{thebibliography}{10}
\expandafter\ifx\csname url\endcsname\relax
  \def\url#1{\texttt{#1}}\fi
\expandafter\ifx\csname urlprefix\endcsname\relax\def\urlprefix{URL }\fi
\providecommand{\bibinfo}[2]{#2}
\providecommand{\eprint}[2][]{\url{#2}}

\bibitem{Cheng2013}
\bibinfo{author}{Cheng, H.} \emph{et~al.}
\newblock \bibinfo{title}{Atomic-level engineering of phase change material for
  novel fast-switching and high-endurance pcm for storage class memory
  application}.
\newblock \emph{\bibinfo{journal}{Electron Devices Meeting (IEDM), 2013 IEEE
  International}} \bibinfo{pages}{30.6.1--30.6.4} (\bibinfo{year}{2013}).

\bibitem{Simpson2011nn}
\bibinfo{author}{Simpson, R.~E.} \emph{et~al.}
\newblock \bibinfo{title}{{Interfacial phase-change memory}}.
\newblock \emph{\bibinfo{journal}{Nature Nanotech.}}
  \textbf{\bibinfo{volume}{6}}, \bibinfo{pages}{501--505}
  (\bibinfo{year}{2011}).

\bibitem{RaouxPCM2009}
\bibinfo{author}{Raoux, S.} \& \bibinfo{author}{Wuttig, M.}
\newblock \emph{\bibinfo{title}{Phase-change materials: Science and
  Applications}} (\bibinfo{publisher}{Springer}, \bibinfo{address}{New York},
  \bibinfo{year}{2009}).

\bibitem{Wuttig2007nmat}
\bibinfo{author}{{Wuttig}, M.} \& \bibinfo{author}{{Yamada}, N.}
\newblock \bibinfo{title}{{Phase-change materials for rewriteable data
  storage}}.
\newblock \emph{\bibinfo{journal}{Nature Mater.}} \textbf{\bibinfo{volume}{6}},
  \bibinfo{pages}{824--832} (\bibinfo{year}{2007}).

\bibitem{shportko2008nm}
\bibinfo{author}{Shportko, K.} \emph{et~al.}
\newblock \bibinfo{title}{Resonant bonding in crystalline phase-change
  materials}.
\newblock \emph{\bibinfo{journal}{Nature Mater.}} \textbf{\bibinfo{volume}{7}},
  \bibinfo{pages}{653--658} (\bibinfo{year}{2008}).

\bibitem{Kolobov2012pss}
\bibinfo{author}{Kolobov, A.~V.}, \bibinfo{author}{Fons, P.},
  \bibinfo{author}{Krbal, M.} \& \bibinfo{author}{Tominaga, J.}
\newblock \bibinfo{title}{Amorphous phase of {GeTe}-based phase-change memory
  alloys: Polyvalency of {Ge-Te} bonding and polyamorphism}.
\newblock \emph{\bibinfo{journal}{Phys. Stat. Sol. (a)}}
  \textbf{\bibinfo{volume}{209}}, \bibinfo{pages}{1031--1035}
  (\bibinfo{year}{2012}).

\bibitem{krbal2012prb}
\bibinfo{author}{Krbal, M.} \emph{et~al.}
\newblock \bibinfo{title}{Crystalline {GeTe}-based phase-change alloys:
  Disorder in order}.
\newblock \emph{\bibinfo{journal}{Phys. Rev. B}} \textbf{\bibinfo{volume}{86}},
  \bibinfo{pages}{045212} (\bibinfo{year}{2012}).

\bibitem{kolobov2014jpc}
\bibinfo{author}{Kolobov, A.~V.}, \bibinfo{author}{Fons, P.},
  \bibinfo{author}{Tominaga, J.} \& \bibinfo{author}{Hase, M.}
\newblock \bibinfo{title}{Excitation-assisted disordering of {GeTe} and related
  solids with resonant bonding}.
\newblock \emph{\bibinfo{journal}{J. Phys. Chem. C}}
  \textbf{\bibinfo{volume}{118}}, \bibinfo{pages}{10248--10253}
  (\bibinfo{year}{2014}).

\bibitem{Li2011prl}
\bibinfo{author}{Li, X.-B.} \emph{et~al.}
\newblock \bibinfo{title}{{Role of Electronic Excitation in the Amorphization
  of Ge-Sb-Te Alloys}}.
\newblock \emph{\bibinfo{journal}{Phys. Rev. Lett.}}
  \textbf{\bibinfo{volume}{{107}}}, \bibinfo{pages}{015501}
  (\bibinfo{year}{{2011}}).

\bibitem{Fons2010prb}
\bibinfo{author}{Fons, P.} \emph{et~al.}
\newblock \bibinfo{title}{Photoassisted amorphization of the phase-change
  memory alloy {$\mathrm{Ge_{2}Sb_{2}Te_{5}}$}}.
\newblock \emph{\bibinfo{journal}{Phys. Rev. B}} \textbf{\bibinfo{volume}{82}},
  \bibinfo{pages}{041203} (\bibinfo{year}{2010}).

\bibitem{Kolobov2011nc}
\bibinfo{author}{Kolobov, A.~V.}, \bibinfo{author}{Krbal, M.},
  \bibinfo{author}{Fons, P.}, \bibinfo{author}{Tominaga, J.} \&
  \bibinfo{author}{Uruga, T.}
\newblock \bibinfo{title}{Distortion-triggered loss of long-range order in
  solids with bonding energy hierarchy}.
\newblock \emph{\bibinfo{journal}{Nature Chem.}} \textbf{\bibinfo{volume}{3}},
  \bibinfo{pages}{311--316} (\bibinfo{year}{2011}).

\bibitem{Hada2015}
\bibinfo{author}{Hada, M.} \emph{et~al.}
\newblock \bibinfo{title}{{Ultrafast time-resolved electron diffraction
  revealing the nonthermal dynamics of near-UV photoexcitation-induced
  amorphization in {$\mathrm{Ge_{2}Sb_{2}Te_{5}}$}}}.
\newblock \emph{\bibinfo{journal}{Sci. Rep.}} \textbf{\bibinfo{volume}{5}},
  \bibinfo{pages}{13530} (\bibinfo{year}{2015}).

\bibitem{Hu2015}
\bibinfo{author}{Hu, J.}, \bibinfo{author}{Vanacore, G.~M.},
  \bibinfo{author}{Yang, Z.}, \bibinfo{author}{Miao, X.} \&
  \bibinfo{author}{Zewail, A.~H.}
\newblock \bibinfo{title}{Transient structures and possible limits of data
  recording in phase-change materials}.
\newblock \emph{\bibinfo{journal}{ACS Nano}} \textbf{\bibinfo{volume}{9}},
  \bibinfo{pages}{6728--6737} (\bibinfo{year}{2015}).

\bibitem{Waldecker2015}
\bibinfo{author}{Waldecker, L.} \emph{et~al.}
\newblock \bibinfo{title}{Time-domain separation of optical properties from
  structural transitions in resonantly bonded materials}.
\newblock \emph{\bibinfo{journal}{Nature Mater.}} \textbf{\bibinfo{volume}{14}},
  \bibinfo{pages}{991--995} (\bibinfo{year}{2015}).

\bibitem{Siders1999}
\bibinfo{author}{Siders, C.~W.} \emph{et~al.}
\newblock \bibinfo{title}{Detection of nonthermal molting by ultrafast x-ray
  diffraction}.
\newblock \emph{\bibinfo{journal}{Science}} \textbf{\bibinfo{volume}{286}},
  \bibinfo{pages}{1340--1342} (\bibinfo{year}{1999}).

\bibitem{Sokolowski-Tinten2003nat}
\bibinfo{author}{Sokolowski-Tinten, K.} \emph{et~al.}
\newblock \bibinfo{title}{Femtosecond x-ray measurement of coherent lattice
  vibrations near the {Lindemann} stability limit}.
\newblock \emph{\bibinfo{journal}{Nature}} \textbf{\bibinfo{volume}{422}},
  \bibinfo{pages}{287--289} (\bibinfo{year}{2003}).

\bibitem{Ichikawa2011}
\bibinfo{author}{Ichikawa, H.} \emph{et~al.}
\newblock \bibinfo{title}{Transient photoinduced `hidden' phase in a
  manganite}.
\newblock \emph{\bibinfo{journal}{Nature Mater.}}
  \textbf{\bibinfo{volume}{10}}, \bibinfo{pages}{101--105}
  (\bibinfo{year}{2011}).

\bibitem{Fons2014}
\bibinfo{author}{Fons, P.} \emph{et~al.}
\newblock \bibinfo{title}{Picosecond strain dynamics in
  {${\mathrm{Ge}}_{2}{\mathrm{Sb}}_{2}{\mathrm{Te}}_{5}$} monitored by
  time-resolved x-ray diffraction}.
\newblock \emph{\bibinfo{journal}{Phys. Rev. B}} \textbf{\bibinfo{volume}{90}},
  \bibinfo{pages}{094305} (\bibinfo{year}{2014}).

\bibitem{Lahme2014}
\bibinfo{author}{Lahme, S.}, \bibinfo{author}{Kealhofer, C.},
  \bibinfo{author}{Krausz, F.} \& \bibinfo{author}{Baum, P.}
\newblock \bibinfo{title}{Femtosecond single-electron diffraction}.
\newblock \emph{\bibinfo{journal}{Struct. Dyn.}} \textbf{\bibinfo{volume}{1}},
  \bibinfo{pages}{034303} (\bibinfo{year}{2014}).

\bibitem{Cavalleri2000}
\bibinfo{author}{Cavalleri, A.} \emph{et~al.}
\newblock \bibinfo{title}{Anharmonic lattice dynamics in germanium measured
  with ultrafast {X-Ray} diffraction}.
\newblock \emph{\bibinfo{journal}{Phys. Rev. Lett.}}
  \textbf{\bibinfo{volume}{85}}, \bibinfo{pages}{586--589}
  (\bibinfo{year}{2000}).

\bibitem{Downer1986}
\bibinfo{author}{Downer, M.~C.} \& \bibinfo{author}{Shank, C.~V.}
\newblock \bibinfo{title}{Ultrafast heating of silicon on sapphire by
  femtosecond optical pulses}.
\newblock \emph{\bibinfo{journal}{Phys. Rev. Lett.}}
  \textbf{\bibinfo{volume}{56}}, \bibinfo{pages}{761--764}
  (\bibinfo{year}{1986}).

\bibitem{Chin1999}
\bibinfo{author}{Chin, A.~H.} \emph{et~al.}
\newblock \bibinfo{title}{Ultrafast structural dynamics in insb probed by
  time-resolved x-ray diffraction}.
\newblock \emph{\bibinfo{journal}{Phys. Rev. Lett.}}
  \textbf{\bibinfo{volume}{83}}, \bibinfo{pages}{336--339}
  (\bibinfo{year}{1999}).

\bibitem{AlsNielsenXrayPhys2011}
\bibinfo{author}{Als-Nielsen, J.} \& \bibinfo{author}{McMorrow, D.}
\newblock \emph{\bibinfo{title}{Elements of Modern X-ray Physics, 2nd Edition}}
  (\bibinfo{publisher}{Wiley-VCH}, \bibinfo{year}{2011}).

\bibitem{lindenberg2005sci}
\bibinfo{author}{Lindenberg, A.~M.} \emph{et~al.}
\newblock \bibinfo{title}{Atomic-scale visualization of inertial dynamics}.
\newblock \emph{\bibinfo{journal}{Science}} \textbf{\bibinfo{volume}{308}},
  \bibinfo{pages}{392--395} (\bibinfo{year}{2005}).

\bibitem{Siegel2004}
\bibinfo{author}{Siegel, J.}, \bibinfo{author}{Schropp, A.},
  \bibinfo{author}{Solis, J.}, \bibinfo{author}{Alfonso, C.} \&
  \bibinfo{author}{Wuttig, M.}
\newblock \bibinfo{title}{Rewritable phase-change optical recording in
  Ge$_{2}$Sb$_{2}$Te$_{5}$ films induced by picosecond laser pulses}.
\newblock \emph{\bibinfo{journal}{Appl. Phys. Lett.}}
  \textbf{\bibinfo{volume}{84}}, \bibinfo{pages}{2250--2252}
  (\bibinfo{year}{2004}).

\bibitem{Park2008tsf}
\bibinfo{author}{Park, I.-M.} \emph{et~al.}
\newblock \bibinfo{title}{Thermomechanical properties and mechanical stresses
  of Ge$_{2}$Sb$_{2}$Te$_{5}$ films in phase-change random access memory}.
\newblock \emph{\bibinfo{journal}{Thin Solid Films}}
  \textbf{\bibinfo{volume}{517}}, \bibinfo{pages}{848 -- 852}
  (\bibinfo{year}{2008}).

\bibitem{shu2013apl}
\bibinfo{author}{Shu, M.~J.}, \bibinfo{author}{Chatzakis, I.},
  \bibinfo{author}{Kuo, Y.}, \bibinfo{author}{Zalden, P.} \&
  \bibinfo{author}{Lindenberg, A.~M.}
\newblock \bibinfo{title}{Ultrafast sub-threshold photo-induced response in
  crystalline and amorphous {GeSbTe} thin films}.
\newblock \emph{\bibinfo{journal}{Appl. Phys. Lett.}}
  \textbf{\bibinfo{volume}{102}}, \bibinfo{pages}{201903}
  (\bibinfo{year}{2013}).

\bibitem{Thomsen1986}
\bibinfo{author}{Thomsen, C.}, \bibinfo{author}{Grahn, H.~T.},
  \bibinfo{author}{Maris, H.~J.} \& \bibinfo{author}{Tauc, J.}
\newblock \bibinfo{title}{Surface generation and detection of phonons by
  picosecond light pulses}.
\newblock \emph{\bibinfo{journal}{Phys. Rev. B}} \textbf{\bibinfo{volume}{34}},
  \bibinfo{pages}{4129--4138} (\bibinfo{year}{1986}).

\bibitem{lyeo2006apl}
\bibinfo{author}{Lyeo, H.-K.} \emph{et~al.}
\newblock \bibinfo{title}{Thermal conductivity of phase-change material
  Ge$_{2}$Sb$_{2}$Te$_{5}$}.
\newblock \emph{\bibinfo{journal}{Appl. Phys. Lett.}}
  \textbf{\bibinfo{volume}{89}}, \bibinfo{pages}{151904}
  (\bibinfo{year}{2006}).

\bibitem{Schick2013cpc}
\bibinfo{author}{Schick, D.} \emph{et~al.}
\newblock \bibinfo{title}{udkm1dsim - a simulation toolkit for 1d ultrafast
  dynamics in condensed matter}.
\newblock \emph{\bibinfo{journal}{Comp. Phys. Comm.}}
  \textbf{\bibinfo{volume}{185}}, \bibinfo{pages}{651 -- 660}
  (\bibinfo{year}{2014}).

\bibitem{Lee2011prl}
\bibinfo{author}{Lee, T.~H.} \& \bibinfo{author}{Elliott, S.~R.}
\newblock \bibinfo{title}{\textit{Ab~Initio} computer simulation of the early
  stages of crystallization: Application to Ge$_{2}$Sb$_{2}$Te$_{5}$
  phase-change materials}.
\newblock \emph{\bibinfo{journal}{Phys. Rev. Lett.}}
  \textbf{\bibinfo{volume}{107}}, \bibinfo{pages}{145702}
  (\bibinfo{year}{2011}).

\bibitem{Akola2007spt}
\bibinfo{author}{Akola, J.} \& \bibinfo{author}{Jones, R.}
\newblock \bibinfo{title}{Structural phase transitions on the nanoscale: The
  crucial pattern in the phase-change materials Ge$_{2}$Sb$_{2}$Te$_{5}$ and
  {GeTe}}.
\newblock \emph{\bibinfo{journal}{Phys. Rev. B}} \textbf{\bibinfo{volume}{76}},
  \bibinfo{pages}{235201} (\bibinfo{year}{2007}).

\bibitem{caravati2007cta}
\bibinfo{author}{Caravati, S.}, \bibinfo{author}{Bernasconi, M.},
  \bibinfo{author}{{K\"uhne}, T.}, \bibinfo{author}{Krack, M.} \&
  \bibinfo{author}{Parrinello, M.}
\newblock \bibinfo{title}{Coexistence of tetrahedral- and octahedral-like sites
  in amorphous phase change materials}.
\newblock \emph{\bibinfo{journal}{Appl. Phys. Lett.}}
  \textbf{\bibinfo{volume}{91}}, \bibinfo{pages}{171906}
  (\bibinfo{year}{2007}).

\bibitem{Micoulaut2014}
\bibinfo{author}{Micoulaut, M.}, \bibinfo{author}{Gunasekera, K.},
  \bibinfo{author}{Ravindren, S.} \& \bibinfo{author}{Boolchand, P.}
\newblock \bibinfo{title}{Quantitative measure of tetrahedral-$s{p}^{3}$
  geometries in amorphous phase-change alloys}.
\newblock \emph{\bibinfo{journal}{Phys. Rev. B}} \textbf{\bibinfo{volume}{90}},
  \bibinfo{pages}{094207} (\bibinfo{year}{2014}).

\bibitem{Loke2012}
\bibinfo{author}{Loke, D.} \emph{et~al.}
\newblock \bibinfo{title}{Breaking the speed limits of phase-change memory}.
\newblock \emph{\bibinfo{journal}{Science}} \textbf{\bibinfo{volume}{336}},
  \bibinfo{pages}{1566--1569} (\bibinfo{year}{2012}).

\bibitem{Boschker2014nano}
\bibinfo{author}{Boschker, J.~E.} \emph{et~al.}
\newblock \bibinfo{title}{Surface reconstruction-induced coincidence lattice
  formation between two-dimensionally bonded materials and a
  three-dimensionally bonded substrate}.
\newblock \emph{\bibinfo{journal}{Nano Lett.}} \textbf{\bibinfo{volume}{14}},
  \bibinfo{pages}{3534--3538} (\bibinfo{year}{2014}).

\bibitem{Katmis2011}
\bibinfo{author}{Katmis, F.} \emph{et~al.}
\newblock \bibinfo{title}{Insight into the growth and control of single-crystal
  layers of Ge--Sb--Te phase-change material}.
\newblock \emph{\bibinfo{journal}{Cryst. Growth Des.}}
  \textbf{\bibinfo{volume}{11}}, \bibinfo{pages}{4606--4610}
  (\bibinfo{year}{2011}).

\bibitem{Rodenbach2012pss}
\bibinfo{author}{Rodenbach, P.} \emph{et~al.}
\newblock \bibinfo{title}{Epitaxial phase-change materials}.
\newblock \emph{\bibinfo{journal}{Phys. Status Sol. RRL}}
  \textbf{\bibinfo{volume}{6}}, \bibinfo{pages}{415--417}
  (\bibinfo{year}{2012}).

\bibitem{Tono2013}
\bibinfo{author}{Tono, K.} \emph{et~al.}
\newblock \bibinfo{title}{Beamline, experimental stations and photon beam
  diagnostics for the hard x-ray free electron laser of sacla}.
\newblock \emph{\bibinfo{journal}{New J. Phys.}} \textbf{\bibinfo{volume}{15}},
  \bibinfo{pages}{083035} (\bibinfo{year}{2013}).

\bibitem{VASP}
\bibinfo{author}{Hafner, J.}
\newblock \bibinfo{title}{Ab-initio simulations of materials using vasp:
  Density-functional theory and beyond}.
\newblock \emph{\bibinfo{journal}{Journal of Computational Chemistry}}
  \textbf{\bibinfo{volume}{29}}, \bibinfo{pages}{2044--2078}
  (\bibinfo{year}{2008}).

\bibitem{feff8}
\bibinfo{author}{Ankudinov, A.} \& \bibinfo{author}{Rehr, J.}
\newblock \bibinfo{title}{Theory of solid state contributions to the x-ray
  elastic scattering amplitude}.
\newblock \emph{\bibinfo{journal}{Phys. Rev. B}} \textbf{\bibinfo{volume}{62}},
  \bibinfo{pages}{2437} (\bibinfo{year}{2000}).

\end{thebibliography}
\end{document}